\documentclass[12pt]{article}
\usepackage{graphicx}
\usepackage{amssymb,amsmath}

\begin{document}

\def\a{\alpha}
\def\b{\beta}
\def\c{\varepsilon}
\def\d{\delta}
\def\e{\epsilon}
\def\f{\phi}
\def\g{\gamma}
\def\h{\theta}
\def\k{\kappa}
\def\l{\lambda}
\def\m{\mu}
\def\n{\nu}
\def\p{\psi}
\def\q{\partial}
\def\r{\rho}
\def\s{\sigma}
\def\t{\tau}
\def\u{\upsilon}
\def\v{\varphi}
\def\w{\omega}
\def\x{\xi}
\def\y{\eta}
\def\z{\zeta}
\def\D{\Delta}
\def\G{\Gamma}
\def\H{\Theta}
\def\L{\Lambda}
\def\F{\Phi}
\def\P{\Psi}
\def\S{\Sigma}
\def\BR{{\rm Br}}
\def\o{\over}
\def\beq{\begin{eqnarray}}
\def\eeq{\end{eqnarray}}
\newcommand{\nn}{\nonumber \\}
\newcommand{\gsim}{ \mathop{}_{\textstyle \sim}^{\textstyle >} }
\newcommand{\lsim}{ \mathop{}_{\textstyle \sim}^{\textstyle <} }
\newcommand{\vev}[1]{ \left\langle {#1} \right\rangle }
\newcommand{\bra}[1]{ \langle {#1} | }
\newcommand{\ket}[1]{ | {#1} \rangle }
\newcommand{\EV}{ {\rm eV} }
\newcommand{\KEV}{ {\rm keV} }
\newcommand{\MEV}{ {\rm MeV} }
\newcommand{\GEV}{ {\rm GeV} }
\newcommand{\TEV}{ {\rm TeV} }
\def\diag{\mathop{\rm diag}\nolimits}
\def\Spin{\mathop{\rm Spin}}
\def\SO{\mathop{\rm SO}}
\def\O{\mathop{\rm O}}
\def\SU{\mathop{\rm SU}}
\def\U{\mathop{\rm U}}
\def\Sp{\mathop{\rm Sp}}
\def\SL{\mathop{\rm SL}}
\def\tr{\mathop{\rm tr}}

\newcommand{\bear}{\begin{array}}  
\newcommand {\eear}{\end{array}}
\newcommand{\la}{\left\langle}  
\newcommand{\ra}{\right\rangle}
\newcommand{\non}{\nonumber}  
\newcommand{\ds}{\displaystyle}
\newcommand{\red}{\textcolor{red}}
\def\ubl{U(1)$_{\rm B-L}$}
\def\REF#1{(\ref{#1})}
\def\lrf#1#2{ \left(\frac{#1}{#2}\right)}
\def\lrfp#1#2#3{ \left(\frac{#1}{#2} \right)^{#3}}
\def\OG#1{ {\cal O}(#1){\rm\,GeV}}

\def\TODO#1{ {\bf ($\clubsuit$ #1 $\clubsuit$)} }


\baselineskip 0.7cm

\begin{titlepage}

\begin{flushright}
IPMU-13-0068 \\
ICRR-report-651-2012-40\\
\end{flushright}

\vskip 1.35cm
\begin{center} 
{\large \bf 
Muon g-2 and 125 GeV Higgs in Split-Family Supersymmetry
} \\
\vskip 1.2cm

{Masahiro Ibe$^{(a,b)}$, Tsutomu T. Yanagida$^{(b)}$, Norimi Yokozaki$^{(b)}$}

\vskip 0.4cm

{\it
$^{(a)}${\it ICRR, University of Tokyo, Kashiwa, 277-8583, Japan} \\
$^(b)$Kavli Institute for the Physics and Mathematics of the Universe,\\
Todai Institutes for Advanced Study, University of Tokyo,\\
Kashiwa 277-8583, Japan\\
}

\vskip 1.5cm

\abstract{
We discuss the minimal supersymmetric standard model with ``split-family" spectrum where 
the sfermions in the first two generations are in the hundreds GeV to a TeV range while
the sfermions in the third generation are in the range of tens TeV.
With the split-family spectrum,  the deviation of the muon $g-2$
and the observed Higgs boson mass are explained simultaneously. 
It is predicted that the gluino and the squarks in the first two generations 
 are within the reach of the LHC experiments 
in most favored parameter space for the universal gaugino mass, which can be tested 
by searching for events with  missing transverse energy or
events with stable charged massive particles.
We also point out that the split-family scenario can be consistent with the focus point
scenario for the non-universal gaugino masses
where the required $\mu$-term is in the hundreds GeV range.
 }
\end{center}
\end{titlepage}

\setcounter{page}{2}

\section{Introduction}
The new particle with mass at around 125--126\,GeV discovered by the ATLAS 
and CMS collaborations\,\cite{Aad:2012tfa,Chatrchyan:2012ufa} 
is confidently believed to be the Higgs boson of the Standard Model.
If low energy supersymmetry (SUSY) is realized in nature in the minimal form (i.e. the MSSM), 
this Higgs boson mass points to the masses of the superpartners of the 
top quark, i.e. the stops, in the tens to hundreds TeV range\,\cite{OYY}.
The straightforward implication of this result is that the whole SUSY spectrum 
or at least the masses of the whole sfermions are in the multi-TeV range or above.%
\footnote{
The later possibilities have been widely discussed 
in the models of pure gravity mediation\,\cite{PGMs},
the models with strong moduli stabilization\,\cite{StrongModuli},
the minimal split supersymmetry\,\cite{ArkaniHamed:2012gw}, where the 
gauginos (and Higgsinos) are in the TeV range.
These models are motivated not only by the Higgs boson mass, but
also by the successful cosmology, where the models are free from
the infamous Polonyi/moduli problems.
}

The heavy SUSY spectrum, however, leads to a tension with another motivation
of  low energy SUSY:  the longstanding deviation of the
observed muon anomalous magnetic moment ($g-2$) from its Standard Model prediction~\cite{g-2_hagiwara2011, g-2_davier2010}.
With the sleptons,  the Higgsino, the wino, and the bino 
in  ${\mathcal O}(100)$\,GeV, 
low energy SUSY has been expected to resolve the discrepancy~\cite{gm2SUSY}. 
The masses of the SUSY particles in the above straightforward expectation 
are too heavy to explain the deviation.

Therefore, it is imperative to investigate possibilities where the observed
 Higgs boson mass and the deviation of the muon $g-2$ can be explained simultaneously.
For example, it has been shown that both can be achieved in the MSSM or in its simple extensions;
\begin{itemize}
\item The observed Higgs boson mass is obtained
in the models with the SUSY spectrum in the hundreds GeV to a TeV range, 
for example,
by the large $A$-term contribution\,\cite{EIY_gm2}, 
or by the extension of the MSSM with extra matter fields\,\cite{EXT_gm2},
or by the extension with extra gauge interactions\,\cite{Endo:2011gy}.
With rather light SUSY spectrum, the deviation of the muon $g-2$ is explained
by the SUSY contributions as in the conventional expectation.
\item The observed Higgs boson mass is obtained 
by the heavy squark contributions while the sleptons are kept light.
Such a colored/non-colored  sfermion mass splitting are achieved, for example, 
in gauge mediation with a messenger multiplet in the adjoint representation 
of  the $SU(5)$ grand unified gauge group\,\cite{Ibe:2012qu}.
With rather light sleptons,  the deviation of the muon $g-2$  gets 
the sizable SUSY contributions.
\end{itemize}

In this paper, we discuss yet another simple possibility, the models with 
the light sfermions in the first two generations and the heavy sfermions in the 
third generation.
Under this simple assumption, the deviation of the muon $g-2$
can be explained due to the light smuons, while the observed Higgs boson mass
is explained by the heavy stop masses.
As we will see, in the case of the universal gaugino mass,  
most of the favored parameter space for the muon anomalous magnetic moment 
can be tested at the LHC experiments 
due to rather light squarks in the first two generations.
We also show that the split-family scenario can be consistent with the focus point
scenario for the non-universal gaugino masses
where the Higgsino is predicted to be in the hundreds GeV range 
in addition to the sleptons in the first two generations.

\section{Split-Family SUSY and the Muon $g-2$}
\subsection*{The muon anomalous magnetic moment}
The muon anomalous magnetic moment, $g-2$, has been measured quite precisely,
which provides us an important probe of new physics beyond the Standard Model.
The experimental value of the muon $g-2$ 
is\,\cite{Bennett:2006fi}:
\begin{eqnarray}
a^{\rm exp}_\mu = 11659208.9(6.3)\times 10^{-10}\ ,
\end{eqnarray}
where $a_\mu = (g-2)/2$.
The prediction of the Standard Model prediction is, on the other hand,
given by \cite{g-2_hagiwara2011};
\begin{eqnarray}
a_\mu^{\rm SM}= 11659182.8(4.9)\times 10^{-10}\ ,
\end{eqnarray}
in which the updated data from  $e^+e^-\to $\,hadrons and the
latest evaluation of the hadronic light-by-light scattering contributions are included.
Therefore, the experimental value of the muon $g-2$ significantly deviates from the Standard Model
prediction more than $3\sigma$, i.e.
\begin{eqnarray}
 \delta a_\mu = a_\mu^{\rm exp} - a_\mu^{\rm SM} = (26.1\pm 8.0)\times 10^{-10}\ .
\end{eqnarray}

In the MSSM, the deviation can be explained  by the SUSY contributions 
when the sleptons (especialfly smuons) as well as the Higgsino and/or the wino and the bino 
are in  ${\mathcal O}(100)$\,GeV. 
(For  precise expressions of the SUSY contributions to the muon
$g-2$, see Ref.\,\cite{Cho:2011rk}.)
As we will see below, the SUSY contributions can be sizable 
while the observed Higgs boson mass is explained simultaneously
in the models with light sfermions in the first two generations
and the heavy third generation sfermions.

\subsection*{Split-family SUSY}
Now, let us introduce  a ``split-family" SUSY scenario, in which the masses of the
sfermions in the first two generations are in ${\mathcal O}(100-1000)$\,GeV
while the masses of the third generation sferimions are in ${\mathcal O}(10)$\,TeV.
In addition, we also assume that the bino and/or wino are also required to be in the hundreds GeV range
since otherwise the SUSY contributions to the muon $g-2$ are suppressed.
The Higgs soft masses are, on the other hand, either as small as the 
first two generation sfermion masses or as large as the third generation masses. 
(We denote  the soft squared masses of the up-type and the down-type Higgs doublets
by $m_{H_u}^2$ and $m_{H_d}^2$, respectively.)
In the following, we simply assume that the sfermion masses in the first two generations
take a universal value $m_0$ at the scale of the grand unified theory (GUT), 
$M_{\rm GUT} \simeq 10^{16}$\,GeV.
The mass of the third generation sfermons are denoted by $m_3$.%
\footnote{
The diagonal split-family soft masses are defined in the basis
where the Yukawa coupling constants take the form of $Y_u = ({Y}_u)_{\rm diag}, \ Y_d = V_{\rm CKM}^* 
({Y}_d)_{\rm diag}$ or $Y_u= V_{\rm CKM}^T ({Y}_u)_{\rm diag}, \ Y_d =  (\hat{Y}_d)_{\rm diag}$
with $V_{\rm CKM}$  being the CKM matrix.
Under this definition, the flavor violating masses which mix the first and second generation squarks 
are given by
\begin{eqnarray}
(\delta_{12}^{d})_{LL} \sim (V_{td}^* V_{ts}) m_{\rm heavy}^2/m_{\rm light}^2\ ,
\end{eqnarray}
where $m_{\rm light}$ and  $m_{\rm heavy}$
are average squark masses in the first two generations and in the third generation, respectively .
(In the latter case, the flavor off-diagonal mass is generated radiatively.) 
The constraint from the $K$--$\bar{K}$ mixing ($\Delta M_K$) can be avoided if  $m_{\rm light}$ is larger than about 1\,TeV\,\cite{FV}.} 

In our numerical analysis, we used {\tt SuSpect} package\,\cite{suspect} to obtain the SUSY spectrum, 
which is modified to include threshold corrections to the slepton masses from the bino, the wino,
the Higgsino, and the heavy Higgs bosons. 
These corrections are included by solving one-loop renormalization group equations below the decoupling scale of the third generation SUSY particles\,\cite{threshold_RGE}.

\subsection*{Case I : Small Higgs soft mass squared}
First, we consider the case where the Higgs soft masses are as small as the sfermion masses
in the first two generations, i.e. $m_{H_u}^2=m_{H_d}^2=m_0^2$ at the GUT scale.
In this case, the Higgs soft masses at the low energy renormalization scale
are dominated by the large radiative corrections from the  third generation masses 
via the renormalization group running.
Thus, the Higgsino mass parameter $\mu$ is predicted to be rather large 
in this case, which is determined by one of the 
minimization conditions of the Higgs potential,
\begin{eqnarray}
\label{eq:EW}
\mu^2 \simeq \frac{m_{H_d}^2 - m_{H_u}^2 \tan^2\b}{\tan^2\b-1} + \frac{1}{2} m_Z^2 
\simeq  - m_{H_u}^2 \ .
\end{eqnarray}
Here, $\tan\b$ is the ratio of the two Higgs vacuum expectation values, 
and we have assumed $\tan\b \gg 1$ in the final expression of Eq.\,(\ref{eq:EW}).
Numerically, we find that the $\mu$-parameter is as large as $8$--$9$\,TeV 
in most parameter space of our interest.
Therefore,  the SUSY contributions to the muon $g-2$ 
is dominated by the one-loop bino-smuon diagram and the one in which the Higgsinos 
are circulating is suppressed\,\cite{cho_hagiwara}.

\begin{figure}[ht]
\includegraphics[scale=1.1]{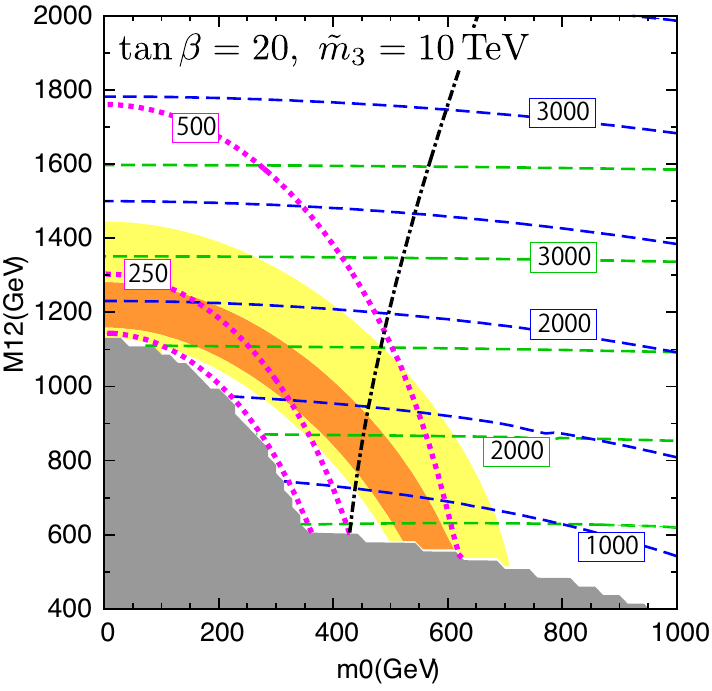}
\includegraphics[scale=1.1]{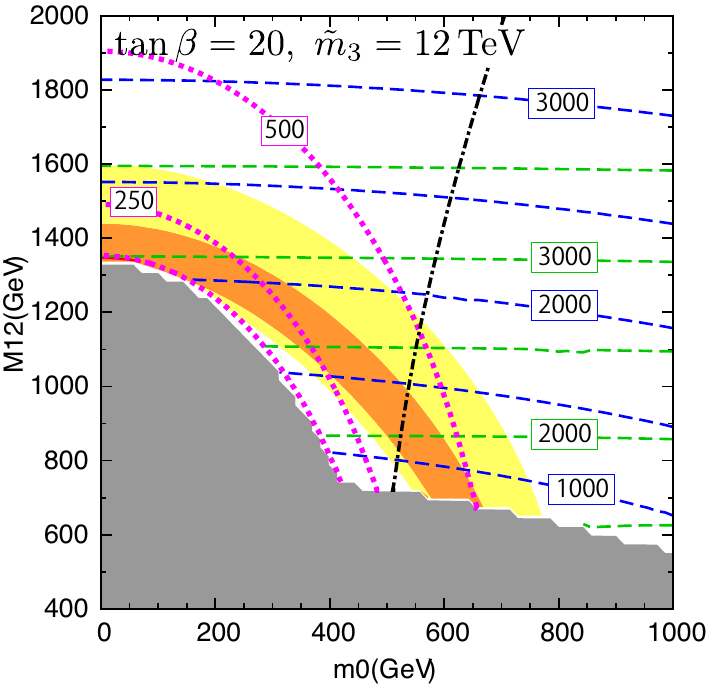}
\includegraphics[scale=1.1]{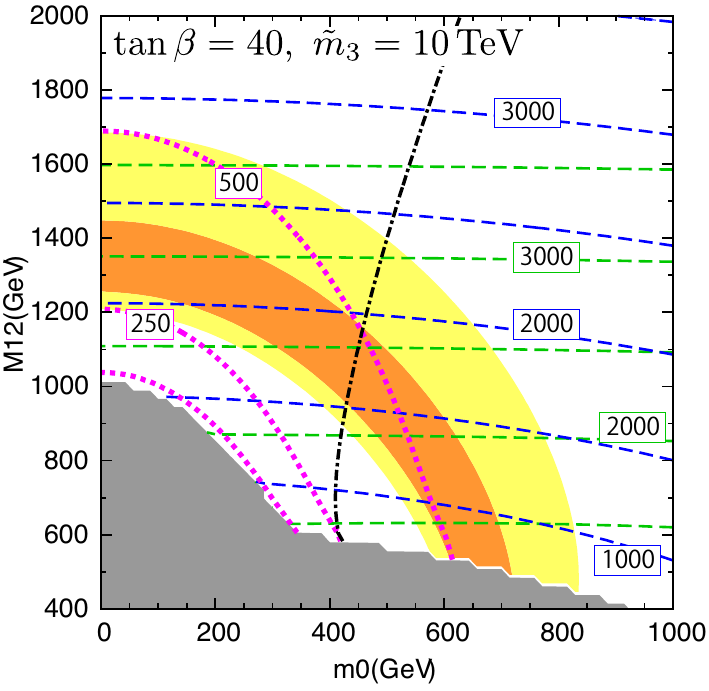}
\includegraphics[scale=1.1]{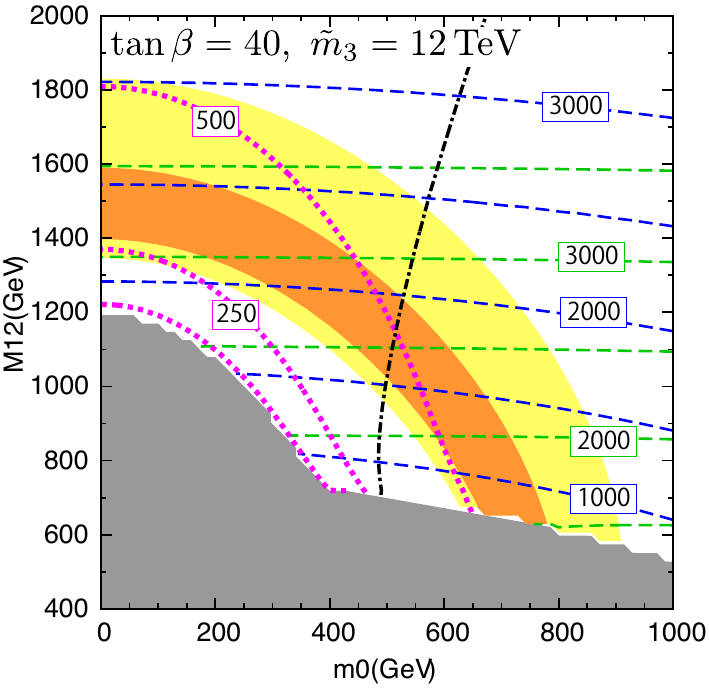}
\caption{\sl\small
Contours of $\delta a_{\mu}$, the squark mass, the gluino mass, and the lightest slepton mass 
(the masses are shown in the unit of GeV) on $m_0-M_{1/2}$ plane. The blue (green) dash-lines correspond to the squark (gluino) masses. The magenta dotted lines show the contours of the lightest slepton masses (from top to bottom, 500 GeV, 250\,GeV, 100\,GeV). In the orange (yellow) region, $\delta a_\mu$ is explained within 1\,$\sigma$ (2\,$\sigma$) level.  On the left region of the black dot-dashed line, the LSP is a slepton. The stop mass is $\simeq 8.5\,(10)$\,TeV for $m_3 = 10\,(12)$\,TeV. 
}
\label{fig:cont1}
\end{figure}

In Fig.\,\ref{fig:cont1}, we show the contours of $\delta a_\mu$, the gluino mass, the squark mass,
and the lightest slepton mass on  $m_0$--$M_{1/2}$ planes.
Here, $M_{1/2}$ is the universal gaugino mass at the GUT scale. 
In the figure, the third generation mass is fixed to $m_{3} = 10$\,TeV and $m_{3} = 12$\,TeV
as reference values with which the lightest Higgs boson mass is in the range of $125$--$126$\,GeV. In the orange (yellow) region, $\delta a_{\mu}$ is explained within $1\,\sigma$ ($2\,\sigma$) level.
The gray shaded regions are excluded where the sleptons or squarks become tachyonic due to 
the large two-loop renormalization group effects from the third generation masses.
The figure shows that the observed muon $g-2$ can be explained within  $1\sigma$ deviation
in a wide parameter region for $\tan\b = {\mathcal O}(10)$.

In the right region of the black dash-dotted line,  the bino-like neutralino is the lightest SUSY particle (LSP) and can be a dark natter candidate. 
Unfortunately, however, the relic abundance of the neutralino is too large in most region of the parameter space. 
This  problem can be solved, for example, by making the LSP short-lived (less than $\sim 1 s$) 
by tiny R-parity violation, which requires another candidate of dark matter such as axion.
In fact, with the tiny R-parity violation, whole region of the parameter space can be cosmologically 
viable, including 
the left region of the black dash-dotted line where the right-handed slepton 
(smuon) is lighter than the lightest neutralino. 
Another possibility is that, as we will discussed below, the abundance of the lightest neutralino can be small enough by bino-wino coannihilation processes\,\cite{bino_wino}, 
which is possible if we  allow the non-universal gaugino masses at the GUT scale.

Before closing this subsection, let us discuss the collider constraints on the model.
In the left region of the black dash-dotted line, the LSP is a charged slepton
which is stable inside the detectors.
In this case, the stringent constraint comes from the search for stable charged particles
by the LHC experiments at the  $8$\,TeV  run using $18.8\,$fb$^{-1}$\cite{CMS_stau}.
For direct slepton production, the current limit is around  $ m_{\rm slepton} \gtrsim 340\,{\rm GeV} $.%
\footnote{
In the favored parameter space for the muon $g-2$,
the gluino and the squarks in the first two generations are typically lighter than $2$\,TeV.
Thus, the constraint on the slepton mass can be more stringent if we 
consider the slepton production via the cascade decays of the gluino/squarks.
The detailed analysis will be given elsewhere.
}
In the right region of the black dash-dotted line, the LSP is the lightest neutralino.
In this case, the  multi-jets with missing transverse energy search at the LHC
puts lower limits on the gluino/squark masses, $m_{\rm gluino/squark}\gtrsim 1.5$\,GeV
at the  $8$\,TeV run using $5.8\,$fb$^{-1}$\cite{ATLAS}.
Thus, most of the parameter space where the observed the muon $g-2$ is explained 
within the $1\sigma$ level has been excluded by the squark mass constraints for $\tan\b = 20$.
Most of the favored parameter space will be also tested at the early stage of the 13\,TeV run. The mass spectra of some reference points are shown in Table.~\ref{table:mass} for the slepton LSP (left) 
and the neutralino LSP (right).

\begin{table}[t!]
  \begin{center}
    \begin{tabular}{  c | c  }
    $m_0,\, m_3$ & 400\,GeV,\ $10$\,TeV \\
    $M_{1/2}$ & 1000 GeV \\
    $\tan\beta$ & 20 \\
    \hline
\hline    
    $\mu$ & $7.7$\,TeV\\
    $m_{\rm stop}$ & 8.5\,TeV \\
    $\delta a_\mu$ & 2.0$\times10^{-9}$ \\
\hline
    $m_{\rm gluino}$ & 2294 GeV \\
    $m_{\rm squark}  $ & 1613 GeV \\
    $m_{\tilde{e}_L} (m_{\tilde{\mu}_L})$ & 610 GeV\\
    $m_{\tilde{e}_R} (m_{\tilde{\mu}_R})$ & 349 GeV \\
    $m_{\chi_1^0}$ & 414 GeV \\
     $m_{\chi_1^{\pm}}$ & 810 GeV \\
    \end{tabular}
        \begin{tabular}{  c | c  }
    $m_0,\, m_3$ & 600\,GeV,\ $12$\,TeV \\
    $M_{1/2}$ & 1100 GeV \\
    $\tan\beta$ & 40 \\
    \hline
\hline    
    $\mu$ & $9.1$\,TeV\\
    $m_{\rm stop}$ & 10\,TeV \\
    $\delta a_\mu$ & 1.9$\times10^{-9}$ \\
\hline
    $m_{\rm gluino}$ & 2512 GeV \\
    $m_{\rm squark}  $ & 1756 GeV \\
    $m_{\tilde{e}_L} (m_{\tilde{\mu}_L})$ & 747 GeV\\
    $m_{\tilde{e}_R} (m_{\tilde{\mu}_R})$ & 568 GeV \\
    $m_{\chi_1^0}$ & 469 GeV \\
     $m_{\chi_1^{\pm}}$ & 896 GeV \\
    \end{tabular}
    \caption{Sample mass spectra for case I.
    The SUSY contributions to $\delta a_\mu$ is also shown.
     }
  \label{table:mass}
  \end{center}
\end{table}

\subsection*{Case II : Large Higgs soft mass squared}
\begin{figure}[t]
\begin{center}
\includegraphics[scale=1.5]{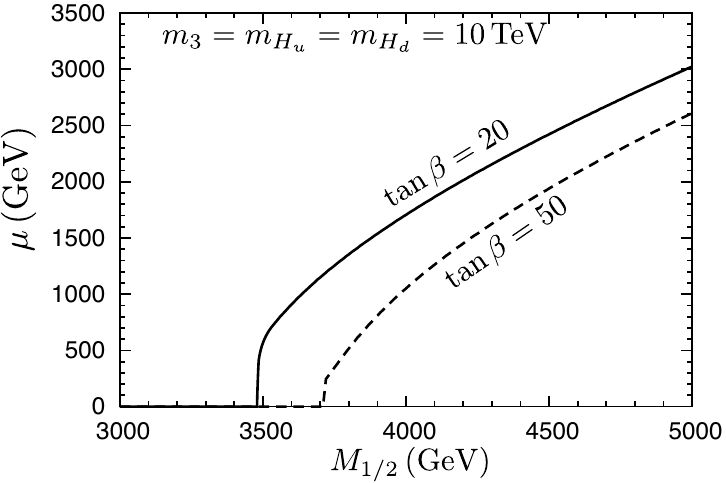}
\caption{\sl \small The Higgsino mass as a function of $M_{1/2}$. Here, we took $m_0=0$\,GeV,
although this assumption is irrelevant for the behavior of the $\mu$-parameter.
The vanishing Higgsino mass for  $M_{1/2}\lesssim 3.5\,(3.7)$\,TeV
for $\tan\b=20 (50)$ signals  unsuccessful electroweak symmetry breaking.
}
\label{fig:noewsb}
\end{center}
\end{figure}

Next, let us consider the case with the Higgs soft masses as large as the 
third generation sfermion masses: $m_{H_u}^2=m_{H_d}^2=m_3^2$. 
In this case, correct electroweak symmetry breaking does not occur when $M_{1/2}$ is
small due to the focus point behavior\,\cite{focus} as shown in Fig.\,\ref{fig:noewsb}. 
In order to realize successful electroweak symmetry breaking, we need $M_{1/2} \gtrsim 3$\,TeV. 
Thus, for the universal gaugino mass at the GUT scale, this  lower limit on $M_{1/2}$
leads to the bino (wino) heavier than 1.2 (2.4)\,TeV.
Furthermore, the large universal gaugino mass  pushes up the slepton masses 
in the first two generations  via the radiative corrections. 
Therefore, in this case, it is difficult to explain the muon $g-2$ deviation as long 
as we assume the universal gaugino mass.

The situation is changed if  the universality of the gaugino masses is relaxed. 
A fascinating example to generate the non-universal gaugino masses 
is a product GUT model\,\cite{PGU}, in which the doublet-triplet splitting problem 
is solved naturally.  
In the product GUT model based on the  $SU(5) \times U(3)_H$ gauge group, 
for example, the gaugino masses at the GUT scale are given by%
\footnote{The factor $(2/5)$ depends on the normalization of $U(1)_H \subset U(3)_H$. 
Here, we take the same normalization as in Ref.~\cite{ibe_watari}.}
\begin{eqnarray}
M_1({\rm GUT}) &\simeq& M_{5} + M_{1H} (2/5) (g_5^2/g_{1H}^2) \, , \nonumber \\
M_2({\rm GUT}) &\simeq& M_{5} \, , \nonumber \\
M_3({\rm GUT}) &\simeq& M_{5} + M_{3H} (g_5^2/g_{3H}^2) \, ,
\end{eqnarray}
where $M_{5}$ denotes the $SU(5)$ gaugino mass, and $M_{1H}$ ($M_{3H}$) 
are the masses of the $U(1)_H \subset U(3)_H$ ($SU(3)_H \subset U(3)_H$) gauginos.
It should be noted that the gauge coupling unification of the MSSM 
gauge coupling constants is effectively maintained,
when the gauge coupling constants of $U(3)_H$ ($g_{1H}$ and $g_{3H}$) are much larger 
than that of $SU(5)$ ($g_5$).
Thus, in this class of the product GUT model, the non-universal gaugino masses can be
realized in a unification consistent manner.
In the following, we simply parameterize the three MSSM gaugino masses at the GUT scale
as free parameters by $M_{1,2,3}$.

\begin{figure}[t]
\begin{center}
\includegraphics[scale=1.1]{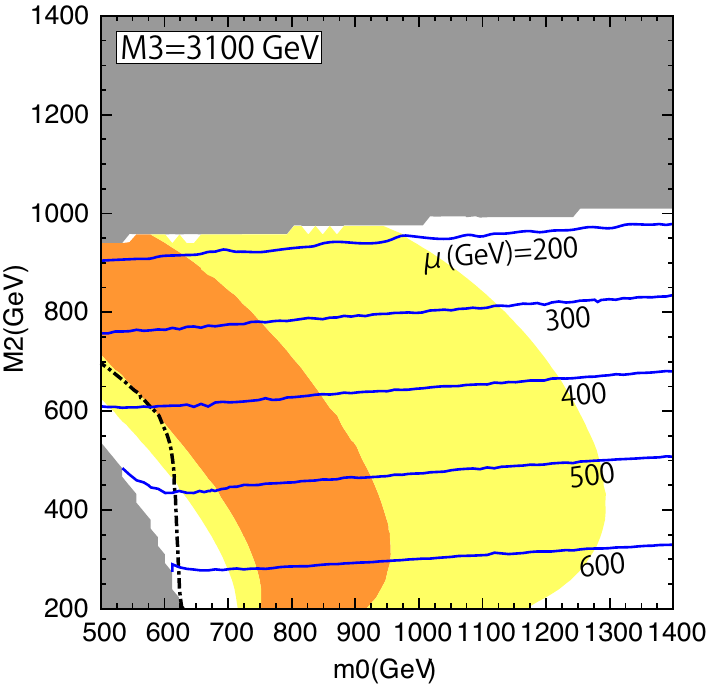}
\includegraphics[scale=1.1]{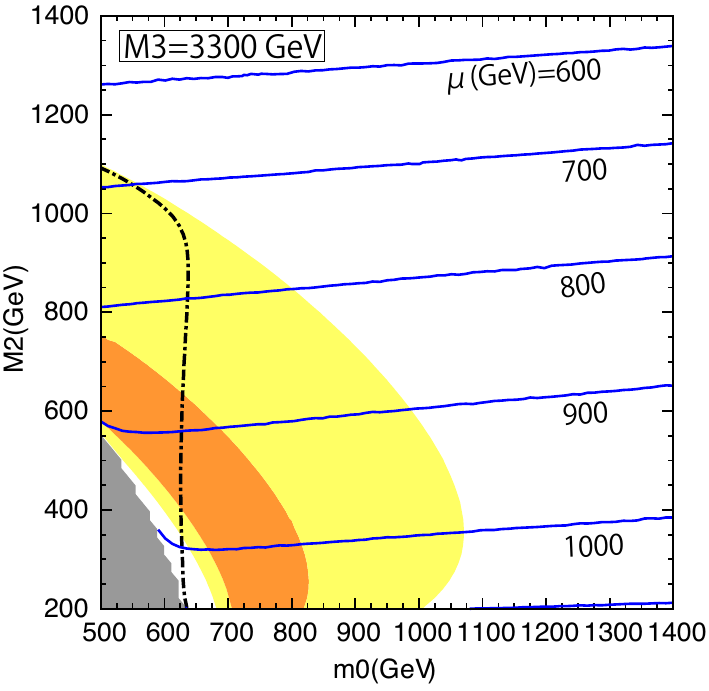}
\includegraphics[scale=1.1]{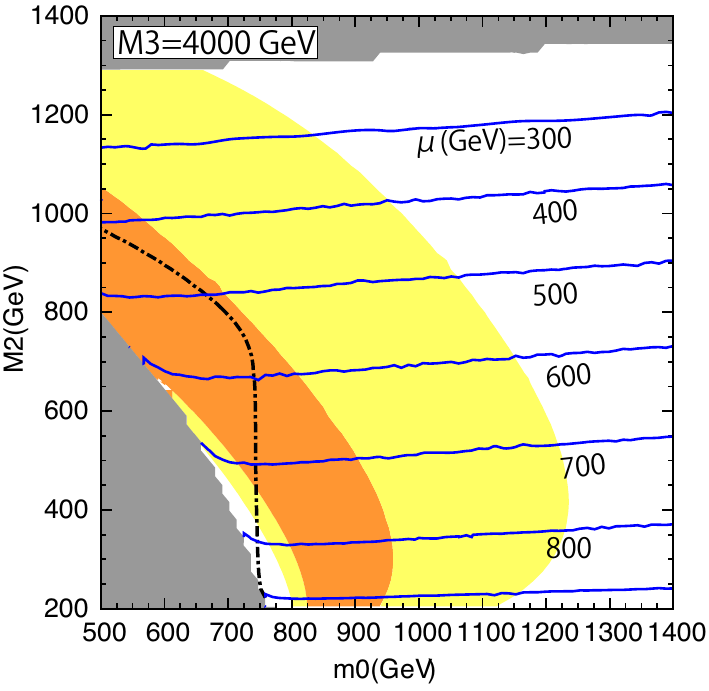}
\includegraphics[scale=1.1]{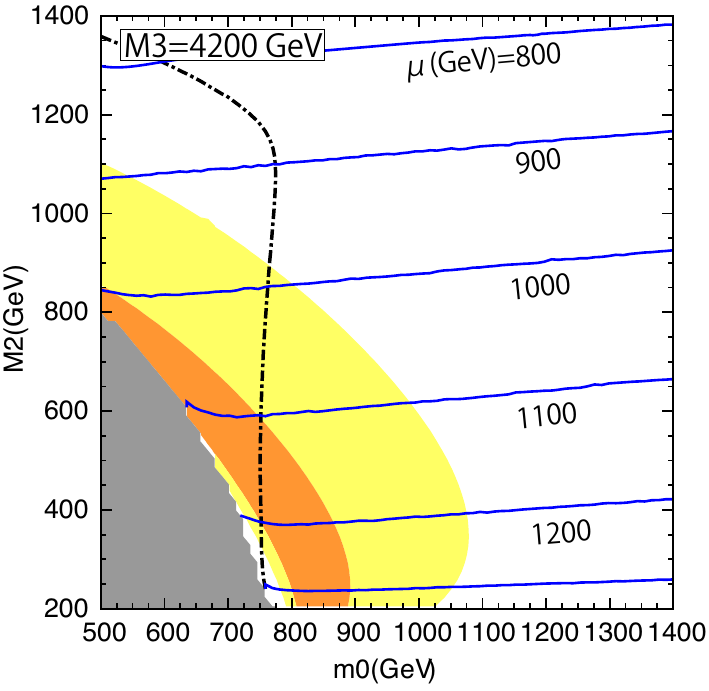}
\caption{\sl \small The contours of the Higgsino mass in the unit of GeV (blue) on $m_0-M_2$ plane for given $M_3$. 
In the orange (yellow) region, $\delta a_\mu$  is explained within 1$\s$ (2$\s$) level. 
Here, $\tan\beta=50$, $M_1/M_2=1.7$. The third generation sfermion mass is taken as $m_3=10$ (12) TeV on upper (lower) two panels. In the orange (yellow) region, $\delta a_\mu$ is explained within 1\,$\sigma$ (2\,$\sigma$) level. The upper gray shaded region is excluded by unsuccessful electroweak symmetry breaking, while the lower gray region is excluded due to the tachyonic slepton. On the left region of the black dot-dashed line, the LSP is a sneutrino.}
\label{fig:higgsino}
\end{center}
\end{figure}

In Fig.~\ref{fig:higgsino}, we show the contours of the Higgsino mass parameter 
$\mu$ and $\delta a_\mu$ on $m_0-M_2$ planes for given $M_3$. 
In the figure, we take $\tan\beta=50$ and $M_1/M_2=1.7$. 
We also take $m_3=10\,(12)$\,TeV for the two top (bottom) planes
so that the lightest Higgs boson mass  is in the range of $125$--$126$\,GeV.
The upper gray region is excluded since electroweak symmetry breaking fails,
near which the $\mu$-parameter is small due to the focus point behavior.
The lower gray region is, on the other hand, excluded due to the tachyonic slepton.
The figure shows that the observed muon $g-2$ can be easily explained at 1$\sigma$ level.
It should be noted that, for $M_1/M_2 = 1.7$, the left-handed sleptons are lighter
than the right-handed sleptons, and hence, the lightest slepton is the sneutrino.   
In the left region of the black dash-dotted line, 
the sneutrino is even lighter than the lightest neutralino.%
\footnote{
The sneutrino LSP dark matter is usually excluded due to the large cross section 
with nuclei\,\cite{sneutrinoDM}, 
unless it is a subdominant component of the dark matter. 
}

In the right region of the black dash-dotted line,
the LSP is the lightest neutralino.
When the neutralino is the bino-like, its relic abundance is too large.
Since we allow the non-universal gaugino mass at the GUT scale, however,
the wino mass can be degenerated with the bino mass when the gaugino
masses satisfy,
\begin{eqnarray}
M_1(M_{\rm GUT}) \sim M_2(M_{\rm GUT}) \frac{\alpha_2(M_{Z})}{\alpha_1(M_{Z})} \simeq 1.8 
M_2(M_{\rm GUT})\ .
\end{eqnarray}
With the degenerated wino/bino masses,  the observed relic dark matter abundance can 
be explained by thermal freeze-out via the wino-bino coannihilations\,\cite{bino_wino}. 

\begin{figure}[t]
\begin{center}
\includegraphics[scale=1.09]{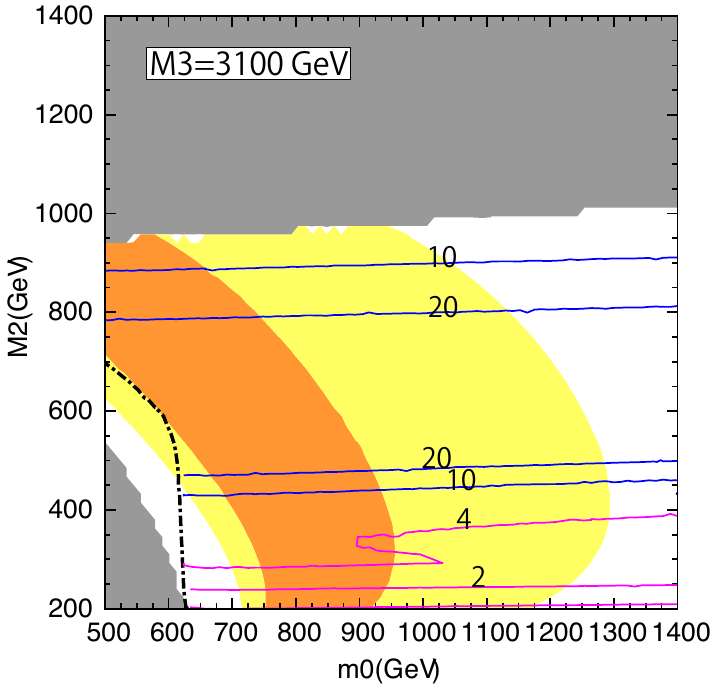}
\includegraphics[scale=1.09]{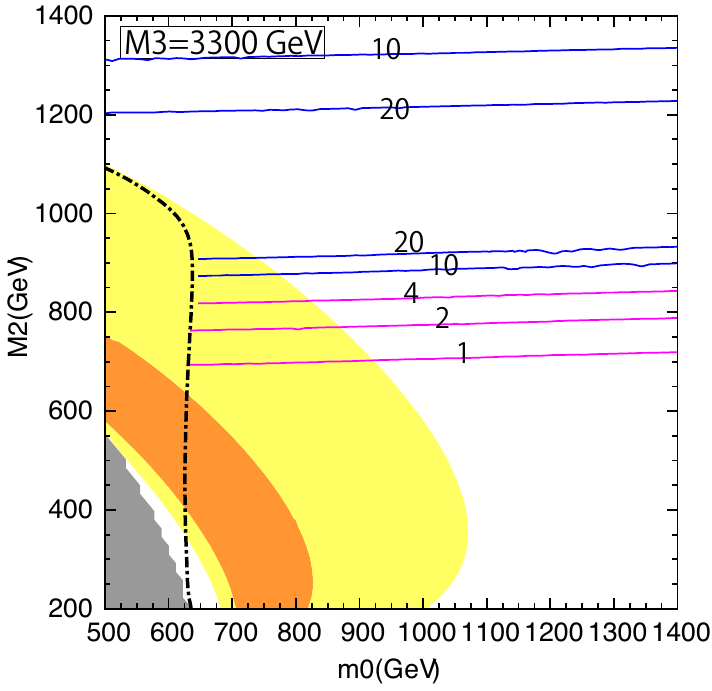}
\includegraphics[scale=1.09]{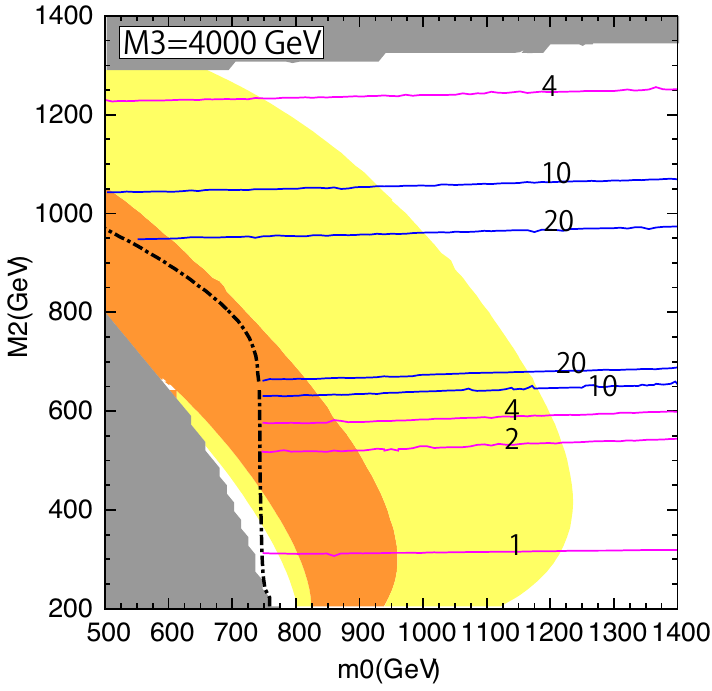}
\includegraphics[scale=1.09]{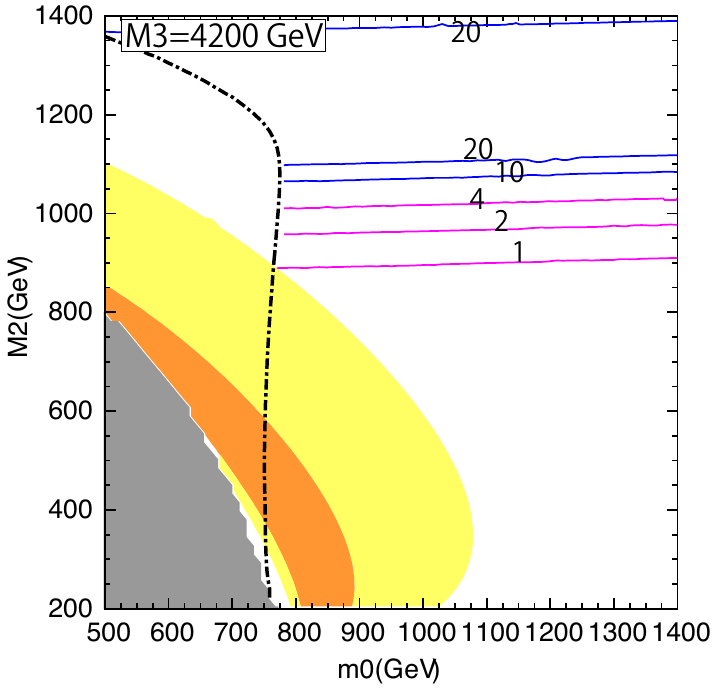}
\caption{\sl \small The spin-independent neutralino-nucleon cross section in the unit of $10^{-45}$\,cm$^2$. Here, $\tan\beta=50$, $M_1/M_2=1.7$ and $f_{s} \simeq 0.26$. ($f_s \equiv \left<N| m_s  \bar{s} s |N\right> / m_N)$ . The third generation scalar mass is taken as $m_3=10\,(12)$ TeV for upper (lower) two panels.}
\label{fig:direct1}
\end{center}
\end{figure}

\begin{figure}[t]
\begin{center}
\includegraphics[scale=1.09]{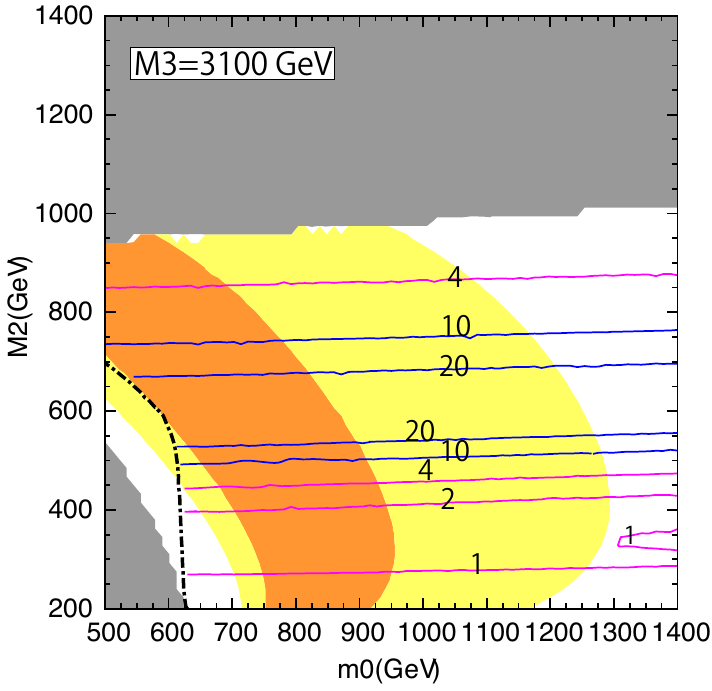}
\includegraphics[scale=1.09]{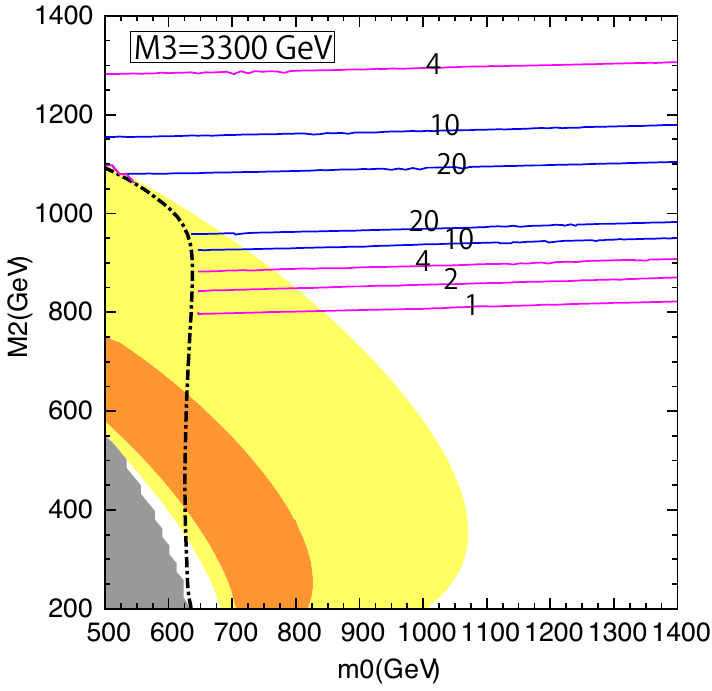}
\includegraphics[scale=1.09]{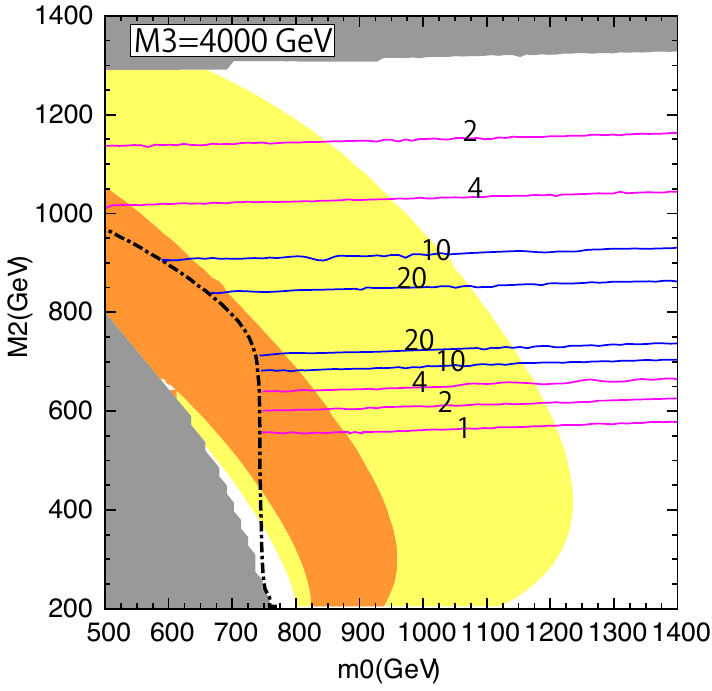}
\includegraphics[scale=1.09]{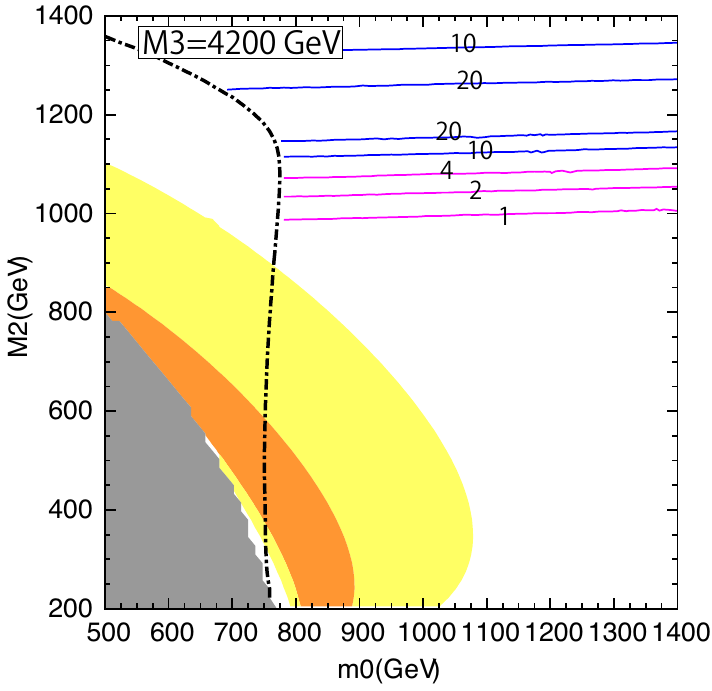}
\caption{\sl\small The spin-independent neutralino-nucleon cross section. Here, $f_{s}=0.009$. Other parameters are same as those in Fig.~\ref{fig:direct1}.}
\label{fig:direct2}
\end{center}
\end{figure}

It should be also noted that the sizable mixing between the Higgsino-gauginos are
expected for $m_{H_u}^2 = m_{H_d}^2 = m_{3}^2$,
since the $\mu$-parameter is  hundreds GeV due to the focus point mechanism.
With the sizable mixing to the Higgsino, the neutralino LSP has 
a large spin-independent cross section to the the neutralino-nucleon,
which is severely constrained by the XENON\,100 experiment\,\cite{Aprile:2012nq};
\begin{eqnarray}
\sigma_p \lesssim 3\times 10^{-45}{\rm cm}^2 \times \left(\frac{m_{\rm DM}}{100\,\rm GeV} \right)\ .
\end{eqnarray} 
(This constraint is valid for the dark matter mass $m_{DM}$ larger than about $100$\,GeV).
In Fig.\,\ref{fig:direct1} and \ref{fig:direct2}, 
we show the contours of the spin-independent cross section.
Here, we again take $\tan\beta=50$ and $M_1({\rm GUT})/M_2({\rm GUT})=1.7$. 
The calculation is performed using {\tt micOMEGAs}\,\cite{micro}. 
In Fig.\,\ref{fig:direct1}, 
we used the strange quark content of the nucleon, $f_s\simeq 0.26$,
which is the default value used in {\tt micOMEGAs},
while the smaller value suggested by the recent lattice calculation~\cite{lattice_strange}, $f_{s}=0.009$, is used in  Fig.\,\ref{fig:direct2}. 
The cross section is suppressed by about a factor of 2 in the latter case.
The figure shows that the spin-independent cross section is too large 
for $M_2 \sim \mu$, where the bino-Higgsino mixing is rather large.
As a result, 
we find that the wino/bino are typically lighter than the Higgsino
in the favored parameter space for the muon $g-2$.

Finally, let us consider the collider constraint.
Due to a rather heavy gluino,
the constraints from the LHC are weaker than the previous section.
Furthermore, since the wino and the bino are rather degenerate,
the constraint from the searches of the three leptons and missing transverse energy 
via the direct gaugino production\,\cite{ATLAS2} is not stringent for the neutralino LSP.
In the left region of the dash-dotted region, the LSP is the sneutrino which is stable inside the detector.
Thus, the main signal is expected to be the direct neutralino/chargino production
leading to two-leptons and missing transverse energy of the sneutrinos,
which is less constrained at the LHC (see Ref.\,\cite{Endo:2013bba} for related discussion).%
\footnote{
Search for events of the highly boosted neutralino, chargino and  slepton pair production
by emitting  a high transverse momentum initial state radiation is expected to enhance 
the discovery potential at the LHC and the ILC.
The detailed analysis on the collider signature will be given elsewhere.
}
The mass spectra of the reference points are shown in Table~\ref{table:mass2} for the neutralino LSP and the sneutrino LSP.

\begin{table}[t!]
  \begin{center}
    \begin{tabular}{  c | c  }
    $m_0,\, m_3$ & 900\,GeV,\ $12$\,TeV \\
    $M_1,\, M_2$ & 820\,GeV, 500\,GeV \\
     $M_3$ & 4000\,GeV \\
    $\tan\beta$ & 50 \\
    \hline
\hline    
    $\mu$ & $701$\,GeV\\
    $m_{\rm stop}$ & 9.9\,TeV \\
    $\delta a_\mu$ & 1.8$\times10^{-9}$ \\
    $\Omega h^2$ & 0.09 \\
\hline
    $m_{\rm gluino}$ & 8.2 TeV \\
    $m_{\rm squark}  $ & 6.7 TeV \\
    $m_A$ & 2.5 TeV \\ 
    $m_{\tilde{e}_L} (m_{\tilde{\mu}_L})$ & 614 GeV\\
    $m_{\tilde{e}_R} (m_{\tilde{\mu}_R})$ & 845 GeV \\
    $m_{\chi_1^0}$ & 335 GeV \\
     $m_{\chi_1^{\pm}}$ & 358 GeV \\
    \end{tabular}
    \begin{tabular}{  c | c  }
    $m_0,\, m_3$ & 600\,GeV,\ $12$\,TeV \\
    $M_1,\, M_2$ & 1360\,GeV, 800\,GeV \\
    $M_3$ & 4000\,GeV \\
    $\tan\beta$ & 50 \\
    \hline
\hline    
    $\mu$ & $519$\,GeV\\
    $m_{\rm stop}$ & 9.9\,TeV \\
    $\delta a_\mu$ & 2.3$\times10^{-9}$ \\
    $\Omega h^2$ & $-$ \\
\hline
    $m_{\rm gluino}$ & 8.2 TeV \\
    $m_{\rm squark}  $ & 6.7 TeV \\
    $m_A$ & 2.3 TeV \\ 
    $m_{\tilde{\nu}_L}$, $m_{e_L}$ & 355\,GeV, 363\,GeV\\
    $m_{\tilde{e}_R} (m_{\tilde{\mu}_R})$ & 649\,GeV \\
    $m_{\chi_1^0}$ & 495\,GeV \\
     $m_{\chi_1^{\pm}}$ & 508\,GeV \\
    \end{tabular}
   \caption{Sample mass spectra, $\delta a_\mu$ and the relic density of the lightest neutralino $\Omega h^2$ for case II. 
     }
  \label{table:mass2}
  \end{center}
\end{table}


\section{Conclusions and discussion}
In this paper, we discussed the MSSM with ``split-family" spectrum
where the sfermions in the first two generations are light while
the sfermions in the third generation is heavy.
With the split-family spectrum, both the deviation of the muon $g-2$
and the observed Higgs boson mass can be explained simultaneously.

In the models with the universal gaugino mass, the gluino/squarks
in the first two generations are typically lighter than $2-3$\,TeV when the deviation muon 
$g-2$ is explained at the $1\sigma$ level.
Therefore, most of the favored parameter space will be tested at the early stage of the LHC 13\,TeV run.
For the non-universal gaugino mass,  which is inevitable for $m_{H_u}^2 = m_{H_d}^2 = m_3^2$,
the gluino/squark masses are rather heavy.
In this case, the collider search of the 
SUSY signals relies on the direct productions of the neutralinos, the charginos, and the sleptons,
which are less constrained at the LHC. 
The discovery potential of those signals is expected to be enhanced by searching for for events of the highly
 boosted neutralino, chargino and  slepton pair productions.

Finally, let us comment on  possible origins of the split-family.
As a simple example, one may consider extra-dimensional model at around the Planck scale
in which the third generation sfermions and the SUSY breaking field
reside on the same brane while the sfermions in the first two generations  are
on a separated brane.
With the geometrical separation, the sfermions in the first generations 
can be much lighter than the ones in the third generation.
As another possibility,  the split-family spectrum can be realized in models with gauge mediation if 
the first two generations and the third generation feel 
two different MSSM gauge interactions, $G_{\rm MSSM}^{(1)}$ and $G_{\rm MSSM}^{(2)}$\,\cite{Craig:2012hc}.
The usual MSSM gauge groups are realized as a diagonal subgroup of $G_{\rm MSSM}^{(1)}\times
G_{\rm MSSM}^{(2)}$ after spontaneous breaking at an appropriate scale below the messenger scale.
In this case, the third generation sfermions can be much heavier when the gauge coupling 
constants of $G_{\rm MSSM}^{(2)}$ is much larger than those of  $G_{\rm MSSM}^{(1)}$.%
\footnote{Required split-family can be also obtained 
if the messenger masses of $G_{\rm MSSM}^{(1)}$ 
are much heavier than those of $G_{\rm MSSM}^{(2)}$. 
}
More ambitious possibility is to identify the first and the second generation 
sfermions with the  pseudo-Nambu-Goldstone bosons 
that parameterize a non-compact K\"ahler manifold\,\cite{Kugo:1983ai,Mandal:2010hc}. 
For example, the coset space $E_7/SO(10)\times U(1)^2$ includes two spinor representations 
and a vector representation of $SO(10)$ which can be identified with the sfermions in the first
two generations and the Higgs multiplets, respectively.
With this identification, we can naturally suppress the soft masses of the sfermions in the first 
two generations as well as the ones of the Higgs doublets.

\section*{Acknowledgements}
We thank S.\,Mastumoto for useful discussion.
This work is 
supported by the World Premier International Research Center Initiative (WPI Initiative), MEXT, Japan.
This work is also supported by Grant-in-Aid for Scientific research from the Ministry of Education, 
Science, Sports, and Culture (MEXT), Japan, No. 22244021 (T.T.Y.), No. 24740151 (M.I.). 
The work of NY is supported in part by JSPS Research Fellowships for Young Scientists.


\end{document}